# Experimental Evidence of Thermal Capillary Waves Excitation on a Microsphere Surface

**Abhishek Sureshkumar[1], Georges Perin[1], Julien Lapeyre[2], Rozenn Bernard[2], Kelig Terrien[3], Bertrand Dudoux[3], Adil Haboucha[3], Hélène Ollivier[1], Yannick Dumeige[1], Stéphane Trebaol[1*]**

[1]ENSSAT, UMR 6082-CNRS-Institut FOTON, Université de Rennes, 6 rue de Kerampont, 22300 Lannion, France
[2]INSA Rennes, UMR 6082-CNRS-Institut FOTON, Université de Rennes, F-35000 Rennes, France
[3]Photonics Bretagne, 4 Rue Louis de Broglie, 22300 Lannion, France

**Abstract**
Whispering-gallery-mode (WGM) microsphere resonators have emerged as a versatile platform across various photonic applications. Despite significant progress, their performance at short wavelengths is fundamentally limited by scattering-induced optical losses that restrict achievable quality factors (Q-factor). Although surface roughness has long been recognised as the leading cause of these losses, its physical origin has remained unclear, with current understanding attributing it to unavoidable fabrication imperfections. Here, we show that thermally excited capillary waves are the fundamental source of scattering losses in microsphere cavities. Using high-resolution atomic force microscopy (AFM) combined with rigorous statistical analysis, we quantitatively identify the characteristic signatures of frozen capillary fluctuations at the sub-nanometre level. The experimentally extracted roughness parameters show close agreement with theoretical predictions based on capillary wave theory. These findings fundamentally revise the prevailing interpretation of surface scattering losses and establish thermodynamic fluctuations, rather than fabrication defects, as the limiting roughness mechanism. By identifying frozen capillary waves as the limiting factor, this work opens new pathways for engineering ultra-high-Q microsphere resonators through fabrication management strategies, particularly for visible- and ultraviolet-photonic applications where scattering losses are most severe.

**Introduction**
Whispering-gallery-mode (WGM) microsphere resonators play a pivotal role in modern photonics[1]. By confining light through continuous total internal reflection at curved dielectric interfaces, these structures support optical modes with exceptionally high quality factors and long photon lifetimes[2]. This remarkable light confinement ability makes WGM resonators a flexible platform for diverse applications. They enable low-threshold lasing in rare-earth-doped glass microspheres[3,4], facilitate a variety of nonlinear optical processes such as second-harmonic generation[5], Kerr[6], Raman[7], or Brillouin scattering[8], and advance cutting-edge technologies in biosensing[9], optical gyroscope[10], quantum optics[11], and microwave photonics[12].

The exceptional performance of WGM resonators is fundamentally determined by the intrinsic optical quality factor (Q-factor), which quantifies photon storage time and total optical loss. State-of-the-art silica microspheres achieve Q-factors approaching $3 \times 10^{11}$ in passive cavities[13] and exceeding $10^{12}$ in erbium-doped active systems[14], corresponding to photon lifetimes reaching millisecond scales at near-infrared wavelengths. These remarkable achievements demonstrate the extraordinary potential of WGM architectures in the telecommunications band. However, this exceptional performance degrades dramatically at shorter wavelengths. Specifically, state-of-the-art silica microspheres at 420 nm exhibit Q-factors limited to approximately $10^8$ [15], a reduction of three to four orders of magnitude relative to their near-infrared counterparts, attributed to roughness-induced surface scattering[16]. This pronounced wavelength dependence severely constrains the extension of high-Q resonators to the visible and ultraviolet spectral regions[15], where improved light confinement could lead to significant advances for many applications.

Prevailing research generally attributes this surface roughness to microscopic defects inherent in the microsphere fabrication process[16–19]. However, systematic Q-factor optimisation has been hindered by significant discrepancies in reported roughness values and the absence of a predictive physical model with broad consensus. Consequently, efforts to minimise optical loss remain largely iterative, constrained by an incomplete understanding of the microscopic mechanisms that govern roughness formation.

This work addresses this fundamental challenge by providing the first direct experimental evidence that surface roughness in silica WGM microsphere cavities is primarily determined by frozen capillary waves[20]. Through multi-scale atomic force microscopy (AFM) and rigorous statistical analysis, we establish quantitative agreement between the observed roughness and theoretical predictions for frozen capillary waves. Our findings confirm that these nanoscale fluctuations, frozen during the solidification process, dictate the surface topography of the microspheres. These results fundamentally reshape the understanding of Q-factor limitations; surface roughness is not an unavoidable fabrication artefact but rather a controllable thermodynamic phenomenon. This insight opens pathways to developing ultra-high-Q resonators through fabrication management optimisation, which is particularly promising for short-wavelength applications.

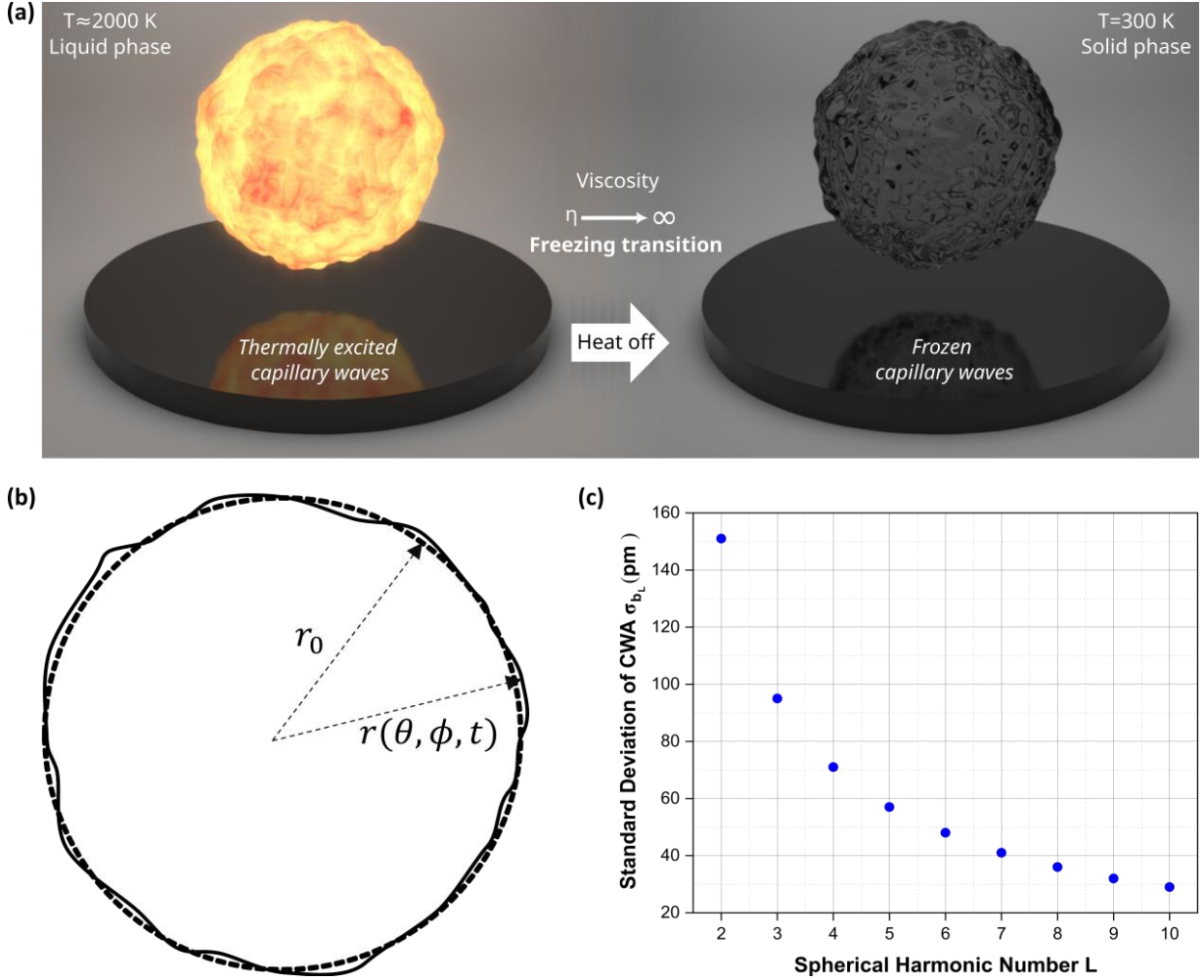

**Fig.1. (a)** Schematic of the thermal capillary waves excitation and freezing process at the surface of a silica glass microsphere. **(b)** Schematic of microsphere surface deformed through thermal capillary wave excitation. **(c)** Theoretical evaluation of the standard deviation of capillary wave amplitude (CWA) as a function of the spherical harmonic index *L*, based on Eq. (3).

Thermal capillary waves represent spontaneous surface fluctuations arising from molecular thermal motion at liquid-vapour or liquid-liquid interfaces[20]. These thermally driven fluctuations exist whenever the surface tension $\gamma$ dominates over gravity, representing a fundamental mechanism through which thermal energy $k_B T$ generates nanoscale surface morphology variations[21]. Height fluctuations related to capillary waves scale as $\sqrt{k_B T/\gamma}$ on the surface plane with commonly reported amplitudes in the hundreds of picometre range. Frozen thermal capillary waves have been experimentally observed in a variety of optical and photonic structures; in particular, their impact on the surface roughness has been experimentally characterised on thin silica films[22], and even on photonic bandgap fibre interfaces[23]. The identification of such a fundamental process further allowed the demonstration of an optimised roughness interface in hollow-core fibre[24], leading to reduced optical propagation losses. These investigations employ statistical analysis techniques[25] to establish a quantitative correspondence between observed roughness and capillary wave theory.

However, no direct surface roughness measurements have yet been reported on a microsphere due to the difficulty in detecting such weak-amplitude surface deformations. During the fabrication process of WGM microsphere cavities, silica rod tips are fused using a heat source[26], generating a molten surface where capillary waves are thermally excited in the liquid phase. The interplay between surface tension and thermal energy determines the amplitude of the capillary waves. As the microsphere cools and solidifies, these surface fluctuations become trapped and frozen into the final surface topology. Gorodetsky *et al.*[16] suggested that the intrinsic surface roughness of WGM microspheres could be determined by such frozen thermal capillary waves. In the present work, we demonstrate experimental evidence that such a framework is appropriate for describing the fundamental surface roughness of glass microspheres.

## Results

### *Frozen capillary waves at the surface of a glass microsphere*
Optical microsphere cavities were fabricated by fusing the tips of pure silica rods using either a fibre fusion splicer or a CO₂ laser as the thermal source[26]. Under continuous heating, the silica rod attains its softening point and undergoes a phase transition to the liquid state (Fig. 1a). Then, thermally excited capillary waves are excited at the surface of the sphere and locally modify the effective radius to[27]:

$$r(\theta,\phi,t) = r_0 + f(\theta,\phi,t) \quad (1)$$

where $r_0$ is the mean radius of the sphere in the absence of thermal excitations (Fig. 1b). The capillary wave amplitude (CWA) can be expressed as:

$$f(\theta,\phi,t) = \sum_{L,M} b_{L,M}(\theta,\phi,t) \times Y_L^M(\theta,\phi) \quad (2)$$

introducing the spherical harmonics decomposition, giving the fundamental basis on which to describe the capillary waves. Each spherical harmonic is weighted by a $b_{L,M}(\theta,\phi,t)$ coefficient that follows a random distribution, driven by thermal fluctuations, characterised by a standard deviation value[28]:

$$\sigma_{b_L} = \sqrt{\frac{k_B T_g}{\gamma(L-1)(L+2)}} \quad (3)$$

where $k_B$ is the Boltzmann constant, $T_g$ is the softening temperature, $\gamma$ is the surface tension, and $L$ is the spherical harmonic index (with $L \neq 0,1$). For the lowest permissible spherical harmonic mode ($L = 2$), the model predicts a standard deviation value of the CWA as $\sigma_{b_2} \approx 151\ pm$ (at softening temperature $T_g \approx 2000 K$[29] and surface tension $\gamma \approx 0.3\ N/m$[30] for silica WGM microsphere cavity). The equipartition principle implies that all accessible capillary wave modes are thermally excited. Each capillary wave can be modelled as a damped harmonic oscillator subjected to a Langevin force proportional to the thermal energy supplied by the heat source[27]. The amplitude of capillary waves as a function of mode number $L$ are shown in Fig. 1c. Since higher-order capillary modes exhibit weaker amplitude, the contribution to the surface deformation is mainly attributed to low $L$ index spherical harmonics.

When the heat source is turned on, and the silica reaches its softening temperature, the molten silica enters the liquid phase (Fig. 1a). The surface capillary waves are thermally excited, and their amplitudes $b_{L,M}(\theta,\phi,t)$ fluctuate along stochastic trajectories described by Gaussian white-noise distributions. The surface state of the sphere, resulting from the linear superposition of capillary waves, undergoes continuous amplitude redistribution at each instant. Upon extinction of the heat source (quenching), the capillary waves freeze in place, resulting in a random, non-deterministic surface state (Fig. 1b). The experimentally observed surface roughness, therefore, results from the superposition of numerous capillary waves, giving the surface a random character.

Since the temperature drops suddenly from 2000 K to 300 K, the rapid increase in viscosity ($\eta$) freezes the microsphere shape with negligible further modification (Fig. 1b). Thus, the spatial average obtained from AFM measurements at various positions on the frozen microsphere can be interpreted as approximating the ensemble average over non-equilibrium configurations of the liquid phase. Specifically, AFM measurements across different positions on the frozen sphere constitute a form of pseudo-temporal averaging, providing access to the same statistical distribution that would be obtained by monitoring a single surface point over extended times during the liquid phase. Thus, the standard deviation of the surface roughness $\sigma_{AFM}$ can be estimated using the procedure described in the next section.

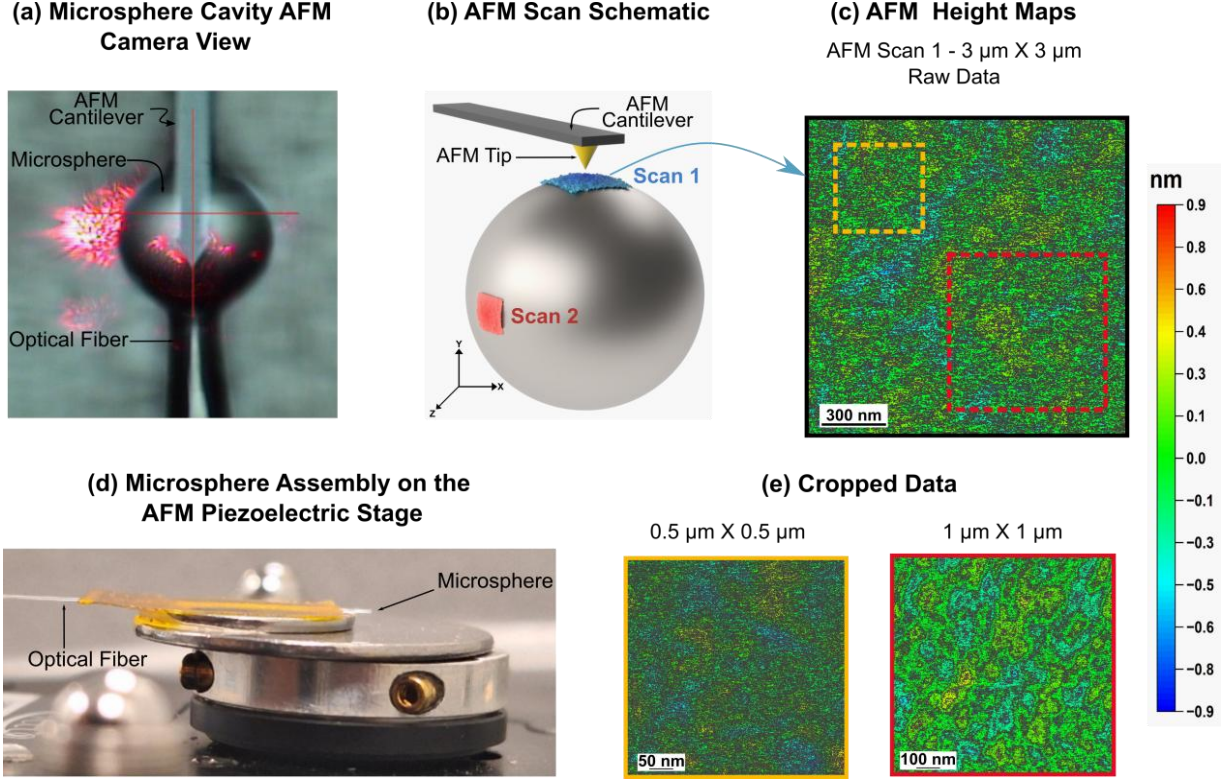

**Fig.2. (a)** Optical camera view of the microsphere attached to an optical fibre during AFM alignment. The AFM cantilever and tip are positioned at the equator and centre of the microsphere to minimise scanning artefacts. **(b)** Schematic illustration of the AFM scanning configuration, showing the AFM tip scanning the surface of the microsphere at different rotational positions to acquire surface topographies. **(c)** Representative raw AFM topography map of microsphere sample S01 acquired over a 3 $\mu m$ × 3 $\mu m$ area, revealing surface height variations at the picometre scale (colour scale in nm). Dashed regions indicate sub-areas (0.5 $\mu m$ × 0.5 $\mu m$ and 1 $\mu m$ × 1 $\mu m$) extracted from the larger dataset. **(d)** Experimental mounting configuration of the microsphere attached to an optical fibre and secured using Kapton tape, with the microsphere suspended in air to ensure mechanical stability and prevent surface deformation during measurements. **(e)** The AFM topography maps corresponding to the cropped regions highlighted in (c), acquired over 0.5 $\mu m$ × 0.5 $\mu m$ and 1 $\mu m$ × 1 $\mu m$ scan areas, respectively, providing detailed insights into nanoscale surface morphology, which were used for subsequent analysis.

## *Tracking frozen capillary waves through AFM cartography*

In this work, surface fluctuations are quantified using two-dimensional (2D) autocorrelation analysis of the AFM surface topography data. AFM offers significant advantages over alternative techniques (e.g., optical profilometry), including exceptional lateral and vertical resolution determined by the AFM tip geometry, combined with non-destructive surface interrogation[31]. Although AFM measurements are time-intensive, the present study extends measurements to substantially larger surface regions compared with previous works, enabling statistically robust and representative surface characterisation. The one-dimensional (1D) autocorrelation function is defined as $R(u_x) = <f(x)f(x+u_x)>$ where $f(x)$ represents the surface height function, $u_x$ is the spatial lag[32] and the angle brackets denote spatial averaging. Extending this to two dimensions (2D), the autocorrelation function becomes $R(u_x, u_y) = <f(x,y)f(x+u_x, y+u_y)>$. A fundamental property of the autocorrelation function is that evaluation at zero spatial lag ($u_x, u_y = 0$) directly yields the standard deviation of roughness amplitude as $\sigma_{AFM} = \sqrt{R(0,0)}$. The evaluation of the roughness at $N$ distinct positions of the microsphere should lead to an evaluation of the average standard deviation of the roughness amplitude through:

$$\langle \sigma_{AFM} \rangle = \sqrt{\frac{1}{N}\sum_{i=1}^{N}\sigma_{AFM}^2(i)} \qquad (4)$$

AFM surface topography measurements were acquired over scan areas ranging from 0.5 $\mu m$ × 0.5 $\mu m$ to 3 $\mu m$ × 3 $\mu m$. Fig. 2 presents a schematic illustration of the WGM microsphere cavity along with representative AFM images demonstrating surface topography acquired at distinct locations on the microsphere surface. Acquired AFM data underwent pre-processing followed by statistical analysis employing the 2D autocorrelation method.

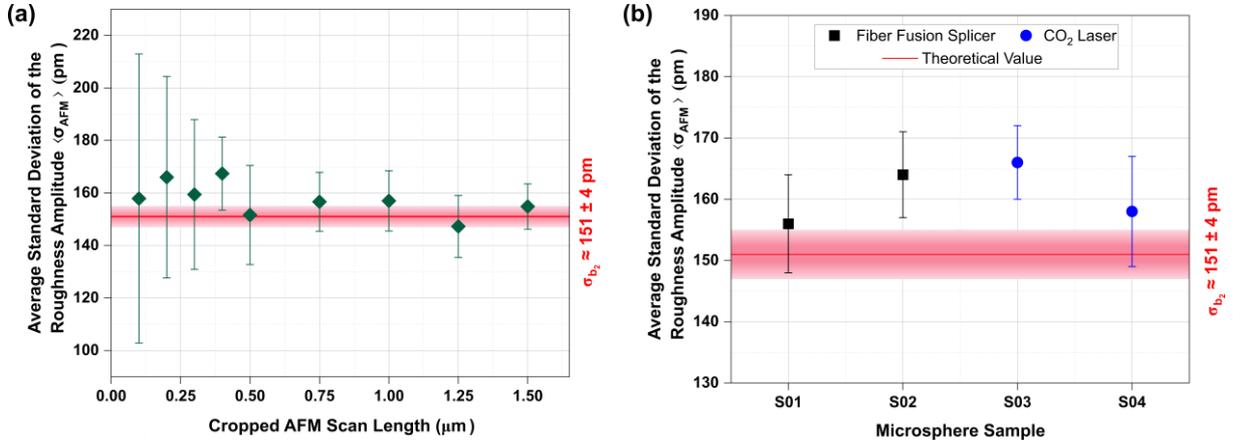

**Fig.3.** Statistical analysis of average standard deviation of the roughness amplitude $\langle \sigma_{AFM} \rangle$ via 2D autocorrelation of AFM Data. **(a)** Dependence of $\langle \sigma_{AFM} \rangle$ as a function of scan size, showing minimal divergence for larger scan areas greater than or equal to 1 $\mu m$ × 1 $\mu m$. This behaviour ensures reliable roughness characterisation independent of scan parameters. Analysis shown for sample S01. **(b)** Experimentally obtained $\langle \sigma_{AFM} \rangle$ for different microsphere samples, evaluated at the largest common scan area of 1.5 $\mu m$ × 1.5 $\mu m$, allowing consistent statistical comparison. Values represent mean ± standard deviation from multiple independent measurements. Reported precision reflects statistical averaging rather than the intrinsic resolution limit of the AFM. The square and circle symbols denote microspheres fabricated using a fibre fusion splicer and a $CO_2$ laser, respectively. The red solid line in both plots indicates the theoretically predicted value $\sigma_{b_2} = 151 \pm 4$ pm, while the shaded region represents the estimated confidence interval resulting from uncertainties in softening temperature $T_g$ and surface tension $\gamma$.

Systematic evaluation of the average standard deviation of the roughness amplitude $\langle \sigma_{AFM} \rangle$ as a function of cropped AFM area was performed and is presented in Fig. 3a. For each selected area size, five independent analyses were performed to evaluate $\langle \sigma_{AFM} \rangle$. This multi-location approach ensures a statistically reliable assessment of the intrinsic surface roughness. The analysis reveals an AFM-size dependence, with $\sigma_{AFM}$ exhibiting a high variability at the sub-micrometre scale due to incomplete evaluation of the surface-height fluctuation amplitude. These observations provide important context for interpreting prior microsphere surface roughness studies, many of which performed analyses without adequate sampling. As a result, $\sigma_{AFM}$ exhibited strong dispersion[19,33] as illustrated in Fig. 3a. Thus, insufficient scan sizes and inadequate sampling potentially lead to systematic under- or over-estimation of $\sigma_{AFM}$.

The average standard deviation for microsphere sample S01 (Fig. 3b) is found to be $\langle \sigma_{AFM} \rangle = 156 \pm 8$ pm, in very good agreement with the theoretical standard deviation of frozen capillary waves, $\sigma_{b_2} = 151 \pm 4$ pm, calculated using Eq. (3). This theoretical value is evaluated at a softening temperature $T_g \approx 1971 \pm 83 \, K$[29] and surface tension $\gamma \approx 0.3 \pm 0.01 \, N/m$[30]. This close correspondence between the experimentally measured surface roughness and the theoretical prediction provides strong evidence that the microsphere surface roughness can be described within the framework of capillary wave theory.

To strengthen the report of frozen capillary wave characterisation, surface roughness analysis was extended to additional microsphere samples. The extracted $\langle \sigma_{AFM} \rangle$ values for each sample, illustrated in Fig. 3b, demonstrate close agreement with theoretical predictions for capillary-wave-induced surface roughness. The present study provides the first experimental evidence supporting the theoretical prediction that frozen capillary waves are the primary source of surface roughness in glass microsphere cavities.

## Discussion

### *Contribution of higher harmonics to the overall surface roughness*

Long-wavelength modes (small $L$) exhibit relatively slow relaxation in molten glass near the softening temperature $T_g$. Upon cooling, these modes tend to freeze before reaching equilibrium, whereas shorter-wavelength modes can partially relax prior to the glass transition. Consequently, the surface roughness is generally dominated by $L = 2$ and relatively depleted at larger L values, although higher modes may still contribute depending on the cooling conditions. This behaviour can be understood in terms of relative mode damping: the $L = 2$ mode experiences minimal damping compared with higher harmonics, which relax more efficiently. As a result, energy is preferentially retained in the lowest capillary wave mode, reinforcing its prominence in the frozen surface landscape, while contributions from $L > 2$ modes are comparatively reduced. These higher-order modes may still provide minor contributions to the overall roughness, arising from the cumulative effect of all modes. Comparison of AFM-measured roughness (cf. Fig. 3b) with estimates from the capillary wave framework considering only $L = 2$ reveals a slight discrepancy, which can be attributed to the damped contributions of higher-order capillary wave modes.

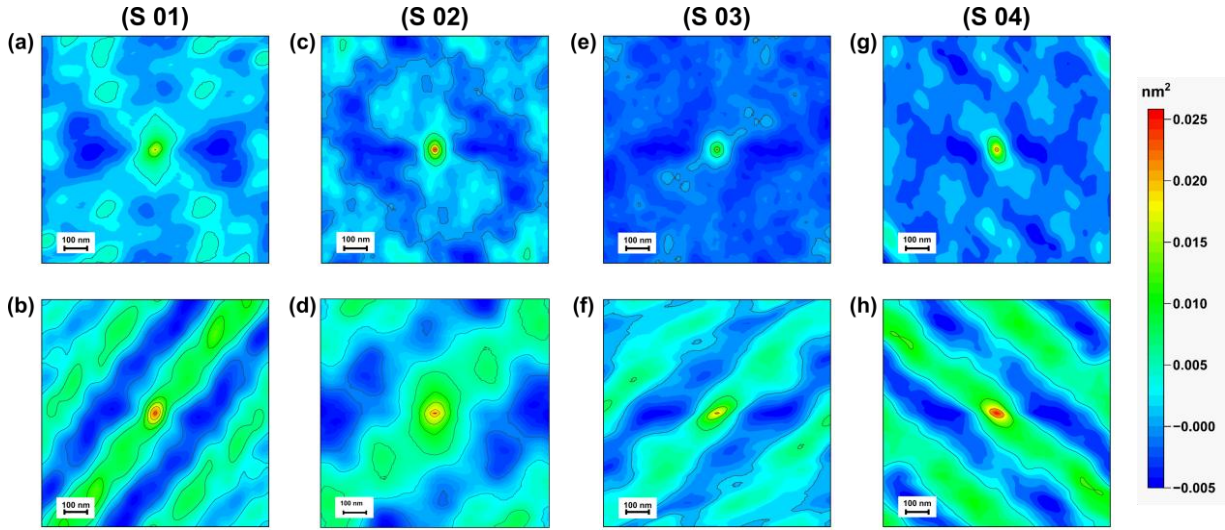

**Fig.4**. 2D-autocorrelation maps of representative microsphere samples. Samples S01 and S02 were fabricated using a fibre fusion splicer, whereas samples S03 and S04 were fabricated using a $CO_2$ laser. The top panels (a, c, e, g) show random-like behaviour observed on the microsphere cavities, while the bottom panels (b, d, f, h) display the anisotropic behaviour on the microsphere. All analyses were performed on AFM scans with an area of $1\,\mu m \times 1\,\mu m$. The anisotropic patterns visible in the autocorrelation maps indicate directional correlations characteristic of frozen capillary waves.

### *Anisotropic nature of microsphere surface texture*

For each sphere, we can analyse different surface texturation revealed through the 2D autocorrelation maps (Fig. 4). Indeed, depending on the position scanned on the microsphere surface, either a random surface (Fig. 4 a, c, e, g) is obtained, or anisotropic behaviours (Fig. 4 b, d, f, h) can also be measured. The heating conditions during microsphere fabrication are globally anisotropic due to the experimental arrangement (electric arc or $CO_2$ laser). Depending on the local orientation of the surface with respect to the heat source and the degree of thermal homogenisation, some regions may experience nearly isotropic relaxation conditions, while others are subjected to more directional thermal gradients. Because the system is quenched through the glass transition on a finite timescale, the surface topography effectively freezes a transient, out-of-equilibrium configuration of the capillary wave landscape, preventing full relaxation towards equilibrium. As a result, the relaxation of capillary wave modes can be locally anisotropic and subsequently imprinted into the surface upon cooling through $T_g$. This naturally accounts for the coexistence, in different regions of the same microsphere, of quasi-isotropic and anisotropic two-dimensional autocorrelation patterns. Similar anisotropic two-dimensional autocorrelation patterns in AFM maps have been reported for silica glass photonic band-gap fibres fabricated under anisotropic conditions[34].

## Conclusion

This study provides the first direct and quantitative experimental evidence that frozen thermal capillary waves constitute the fundamental source of surface roughness in silica WGM microsphere resonators, fundamentally reshaping the understanding of optical loss mechanisms in these cavities. Through comprehensive AFM characterisation combined with rigorous statistical analysis, we demonstrate quantitative agreement between experimentally determined roughness values and theoretical predictions from capillary wave physics. This convergence of theory and experiment across multiple samples and fabrication methods provides compelling evidence for the capillary wave hypothesis.

The implications of this finding are profound; surface roughness is not an unavoidable by-product of glass processing but rather a controllable thermodynamic phenomenon determined by the molten surface properties during fabrication. This realisation opens new optimisation pathways through fabrication management strategies. Since capillary wave amplitude is governed by the surface tension and viscosity, rational manipulation of temperature profiles, cooling rates, and atmospheric conditions during fabrication can suppress capillary wave excitation or accelerate their dissipation, thereby reducing surface roughness[23,35]. Such optimisation strategies may enable dramatic improvements in Q-factor performance, particularly at short wavelengths where surface scattering dominates all loss channels. This work thus establishes a practical pathway toward the generation of optical cavities with performance previously thought inaccessible, with implications for quantum sensing, nonlinear photonics, and optical information processing.

## Methods

### *Sample Fabrication and Characterisation*

WGM microsphere cavities were fabricated based on the formation of spherical structures from melted optical fibres driven by surface tension. High-purity silica rods (F300) were used as the starting material. Two samples S01 ($257~\mu m$) and S02 ($241~\mu m$) were fabricated at the Institute FOTON using a pre-programmed fibre fusion splicer FSU 995. In this process, the polymer coating on the silica rod was first removed, and the rod was thoroughly cleaned using alcohol. It was then heated for several hundred milliseconds before the heating was abruptly terminated, allowing the molten tip to reflow into a spherical shape. Two additional samples, S03 ($177~\mu m$) and S04 ($164~\mu m$) were fabricated at Photonics Bretagne using a $CO_2$ laser integrated Smartsplicer from Nyfors. In this case, the silica rod was heated for several tens of milliseconds, after which the heating was stopped instantaneously, thereby enabling surface-tension-driven sphere formation. A detailed description of the fabrication protocols is described in refs[15,26,36]. Following fabrication, the microsphere samples were optically characterised via cavity ring-down measurements, from which the cavity's intrinsic Q factors were evaluated to be close to $10^8$ [37].

### *Data acquisition procedure using AFM*

The roughness of the microsphere surface are characterised using AFM in tapping mode (TM-AFM). Measurements are performed with a Bruker Innova AFM, equipped with an Al-coated tip (Nanoworld Arrow-NCR model). The tip, with a radius of less than 10 nm, features an Al coating thickness of 30 nm, a resonant frequency of 285 kHz, and a spring constant of 42 N/m. Images are recorded at scan frequencies ranging from 0.1 to 0.4 Hz, with a resolution of $1024 \times 1024$ pixels.

Given the extremely smooth surface of the studied microsphere, particular attention must be paid to both the analysis conditions and data processing. During initial observations, we noted a slight flattening of the microsphere when its surface rested on the pad. This resulted in a localised alteration of the surface topography at the contact area with the metallic pad. To avoid any modification of the surface profile at the contact point, the microsphere must always remain suspended. Since the microsphere is fused to the tip of an optical fibre, it is mechanically supported by the fibre itself. We secure the microsphere by fixing the optical fibre with Kapton tape to metallic pads of various diameters. The microsphere is then suspended in air, and the area of analysis can be adjusted by rotating the optical fibre (Fig. 2d). Measurements are taken at multiple locations and over varying area sizes ($0.5~\mu m \times 0.5~\mu m$ to $3~\mu m \times 3~\mu m$), with the tip localised at the equator and centre to avoid geometric artefacts. However, the scan area cannot be increased indefinitely because the bowing effect must account for the curvature of the microsphere cavities, so the AFM scans were limited to $3~\mu m \times 3~\mu m$.

*AFM data analysis*

The acquired AFM datasets were pre-processed to remove experimental artefacts and surface imperfections prior to quantitative analysis. To study how the roughness standard deviation ($\sigma_{AFM}$) changes with scan size, the AFM measurements were performed on the surface of the microspheres. Topographical data were collected over scan areas ranging from $0.5~\mu m \times 0.5~\mu m$ up to $3~\mu m \times 3~\mu m$ at different locations on each microsphere. From these larger scans, smaller square regions were systematically extracted, starting with areas of $0.1~\mu m \times 0.1~\mu m$. The size of these cropped regions was then gradually increased, allowing the $\sigma_{AFM}$ to be analysed across multiple length scales. For each scan size, five separate regions were selected from different positions within the AFM maps, as shown in Fig. 2c. The roughness values calculated from these regions were then averaged to obtain a representative $\langle\sigma_{AFM}\rangle$ value for that specific window size. This averaging helps minimise the effect of local surface variations. The resulting relationship between $\langle\sigma_{AFM}\rangle$ and the effective cropped scan size of analysis is shown in Fig. 3a. For the $\langle\sigma_{AFM}\rangle$ values reported in Fig. 3b, a similar analysis procedure was employed. For all samples, a scan area of $1.5~\mu m \times 1.5~\mu m$ was selected from the larger AFM maps. In total, five independent cropped sub-regions were extracted and individually analysed to determine $\langle\sigma_{AFM}\rangle$. The reported values correspond to the average of these five measurements, providing a statistically representative estimate of the surface roughness for each sample.

**Acknowledgements**

The authors acknowledge RENATECH+ within nanoRennes for surface analysis support and the financial support of the Indo-French Centre for the Promotion of Advanced Research (70T12-2, MASSALAQ), I-DEMO PIA4 Bpifrance - Région Bretagne - Lannion Tregor Communauté (40898814/1, QoQeliQo), and Lannion Tregor Communauté - Région Bretagne (ARED) (ELVIS Project).

**Data availability**

The datasets generated and analysed during the current study are available from the corresponding authors upon reasonable request.

**Conflict of interest**

The authors declare no competing interests.